\documentclass{icrc2009}

\usepackage{graphicx}   % for including figures
\usepackage{caption}    % for captions
\usepackage[font=footnotesize]{subfig} % subfig.sty for a double column floating figure using two subfigures
\usepackage{fixltx2e}
\usepackage{url}

\newcommand{\shorttitle}[1]%
{\markboth{Proceedings of the 31\MakeLowercase{$^{st}$} ICRC, {\L}\'{o}d\'{z} 2009}{#1} }
\newcommand{\etal}{\MakeLowercase{\textit{et al. }}} % "et al."
\def\deg{{${}^\circ$}}
\def\epeak{$E_\mathrm{peak}$}

\newcommand{\lesssim}{\lower.5ex\hbox{$\; \buildrel < \over\sim \;$}}
\newcommand{\gtrsim}{\lower.5ex\hbox{$\; \buildrel > \over\sim \;$}}

\begin{document}
\title{{\em Fermi} Observations of high-energy gamma-ray emissions \\from GRB 080916C}

\author{\IEEEauthorblockN{Hiroyasu Tajima\IEEEauthorrefmark{1},
			   for {\em Fermi} LAT and {\em Fermi} GBM collaborations%\IEEEauthorrefmark{2}
			   }
                            \\
\IEEEauthorblockA{\IEEEauthorrefmark{1}Kavli Institute of Particle Astrophysics and Cosmology, SLAC National Accelerator Laboratory, USA}
%\IEEEauthorblockA{\IEEEauthorrefmark{2}something?}
}

% please write the preseter's name and short title (3-4 words maximum)
%    which will appear at the header of the even pages.
\shorttitle{H. Tajima \etal {\em Fermi} observations of GRB 080916C}
\maketitle

%******************************************************
% ABSTRACT
%******************************************************
\begin{abstract}
Observations of the long-duration Gamma-Ray Burst GRB~080916C by the {\em Fermi} Gamma-ray Burst Monitor and Large Area Telescope show that it has a single spectral form from 8~keV to 13.2~GeV. 
The  E$\;>\;$100$\;$MeV emission was $\approx\;$5~s later than the E$\;\lesssim\;$1$\;$MeV emission and lasted much longer even after photons with E$\;<\;$100$\;$MeV became undetectable. 
The redshift from GROND of z$\;\approx\;$4.35  means that this GRB has the largest reported apparent isotropic $\gamma$-ray energy release, $E_\mathrm{iso}\approx 8.8 \times 10^{54}\;$ergs. It also sets a stringent lower limit on the GRB outflow Lorentz factor, $\Gamma_{\rm min} \approx\;$890, and limits the quantum gravity mass scale, $M_{QG} > 1.3 \times 10^{18}\;$GeV$/c^2$.
  \end{abstract}


%******************************************************z
% INTRODUCTION
%******************************************************
\section{Introduction}
Gamma-ray bursts (GRBs) are the most luminous explosions in the universe and are leading candidates for the origin of ultra high-energy cosmic rays (UHECRs). 
Prompt emission from GRBs from $\sim$10~keV to $\sim$1 to 5~MeV has usually been detected but, occasionally photons above 100~MeV have been detected by the Energetic Gamma-Ray Experiment Telescope (EGRET)\cite{Dingus95}, and more recently by Astro-rivelatore Gamma a Immagini LEggero (AGILE)\cite{AGILE08}. 
Observations of gamma rays with energies $>100$~MeV are particularly prescriptive because they constrain the source environment and help understand the underlying energy source. 
Although there have been observations of photons above 100~MeV\cite{Hurley94,Wren02,Gonzalez03}, it has not been possible to distinguish competing interpretations of the emission\cite{Granot03,Katz94,Dermer06}. 
The {\em Fermi} Gamma-ray Space Telescope, launched on 11 June 2008, provides broad energy coverage and high GRB sensitivities through the Gamma-ray Burst Monitor (GBM) and the Large Area Telescope (LAT)\cite{LAT}.
The GBM consists of 12 sodium iodide (NaI) detectors which cover the energy band between 8~keV and 1~MeV, and two bismuth germanate (BGO) scintillators which are for the energy band between 150 keV and 40~MeV. The LAT is a pair conversion telescope with the energy coverage from 20~MeV to more than 300~GeV. In this paper, we report detailed measurements of gamma-ray emission from the GRB 080916C detected by the GBM and LAT.

%******************************************************
% Observation
%******************************************************
\section{Observations}
At 00:12:45.613542 UT ($T_0$) on September 16 2008 the GBM flight software triggered on GRB 080916C. 
The GRB produced large signals in 9 of the 12 NaI detectors and in one of the two BGO detectors. Analysis of the data on the ground localized the burst to a Right Ascension (RA) = $08^h07^m12^s$, Declination (Dec) = $-61^\circ18^\prime00^{\prime\prime}$\cite{GCN8245}, with an uncertainty of 2.8\deg\ at 68\% confidence level (C.L.).
At the time of the trigger, the GRB was located $\sim$48\deg\ from the LAT boresight and on-ground analysis revealed a bright source consistent with the GRB location. Using the events collected during the first 66 s after $T_0$, within 20\deg\ around the GBM burst position, the LAT provided a localization of RA = $07^h59^m31^s$, Dec. = $-56^\circ35^\prime24^{\prime\prime}$\cite{GCN8246} with a statistical uncertainty of 0.09\deg\ at 68\% C.L. (0.13\deg\ at 90\% C.L.) and a systematic uncertainty smaller than $\sim$0.1\deg.

Follow-up X-ray and optical observations revealed a fading source at RA = $07^h59^m23.24^s$, Dec. = $-56^\circ38^\prime16.8^{\prime\prime}$ ($\pm1.9^{\prime\prime}$ at 90\% C.L.)\cite{GCN8261} by Swift/X-Ray Telescope (XRT) and RA = $07^h59^m23.32^s$, Dec. = $-56^\circ38^\prime18.0^{\prime\prime}$ ($\pm0.5^{\prime\prime}$)\cite{GCN8257,GCN8272} by Gamma-Ray Burst Optical/Near-Infrared Detector (GROND), respectively, consistent with the LAT localization within the estimated uncertainties. 
GROND determined the redshift of this source to be $z=4.35\pm0.15$\cite{greiner09}.
The afterglow was also observed in the near-infrared band by the Nagoya-SAAO 1.4~m telescope (IRSF)\cite{GCN8274}. 
%The X-ray lightcurve of the afterglow from $T_0+61$~ks to $T_0+1306$~ks shows two temporal breaks at about 2 and 4 days after the trigger\cite{GCN166.1}.
%The lightcurves before, between and after the breaks can be fit with a power-law function with decay indices $\sim-2.3$, $\sim-0.2$ and $\sim-1.4$, respectively.

\begin{figure}[tbhp]
\begin{center}
\includegraphics[width=0.9\linewidth]{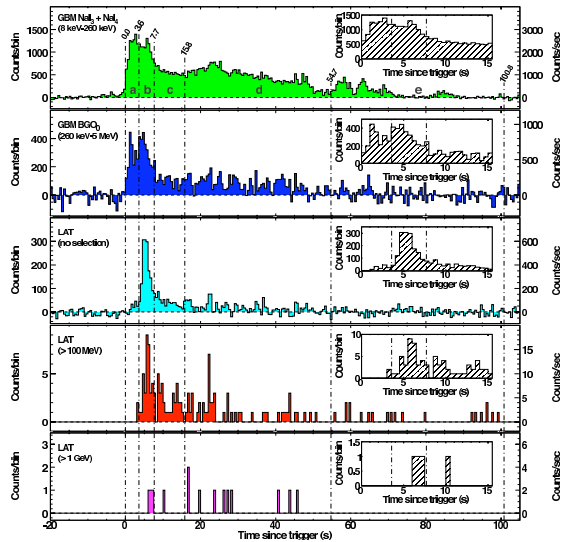}
\caption{Lightcurves for GRB 080916C observed by the GBM and the LAT,
from lowest to highest energies.
The energy ranges for the top two panels are chosen to avoid overlap.
The top three panels represent the background-subtracted lightcurves for the
NaI, the BGO and the LAT.
The top panel shows the sum of the counts, in the~8--260~keV energy band, of two NaI detectors (3 and 4). %\del{with the highest signal.}
The second is the corresponding plot for BGO detector 0, between 260~keV and 5~MeV.
The inset panels give views of the first 15~s from the
trigger time. In all cases, the bin width is 0.5~s; the per-second
counting rate is reported on the right for convenience.} 
\label{fig:GBM-LAT-LC}
\end{center}
\end{figure}

The lightcurve of GRB 080916C, as observed with {\em Fermi} GBM and LAT, is shown in Fig.~\ref{fig:GBM-LAT-LC}. 
The total number of LAT counts after background subtraction in the first 100~s after the trigger was $> 3000$. 
For most of the low-energy events, however, extracting reliable directional and energy information was not possible. After we applied standard selection cuts\cite{LAT} for transient sources with energies greater than 100~MeV and directions compatible with the burst location, 145 events remained (panel 4), and 14 events had energies $> 1$~GeV.

Because of the energy-dependent temporal structure of the lightcurve, we divided the lightcurve into five time intervals (a,b,c,d,e) delineated by the vertical lines (Fig.~\ref{fig:GBM-LAT-LC}). 
The GRB lightcurve at low energy has two bright peaks, one between 0 and 3.6~s after the trigger (interval `a'), and one between 3.6 and 7.7~s (interval `b'). 
The two peaks are distinct in the BGO lightcurve, but less so in the NaI. 
In the LAT detector the first peak is not significant though the lightcurve shows evidence of activity in time interval (a), mostly in events below 100 MeV. Above 100 MeV, peak (b) is prominent in the LAT lightcurve. 
Interval (c) coincides with the tail of the main pulse, and the last two intervals reflect temporal structure in the NaI lightcurve and have been chosen to provide enough statistics in the LAT energy band for spectral analysis.
The highest energy photon was observed during interval (d): $E = 13.22^{+0.70}_{-1.54}$~GeV. Most of the emission in peak (b) shifts toward later times as the energy increases (inset).

%******************************************************
% Spectral analysis
%******************************************************
\section{Spectral Analysis}
We performed simultaneous spectral fits of the GBM and LAT data for each of the five time bins described above and shown in Fig.~\ref{fig:GBM-LAT-LC}.
GBM NaI data from detectors 3 and 4 were selected from 8~keV to 1.0~MeV, as well as BGO detector 0 data from 0.26 to 40~MeV. 
LAT photons were selected using the ``transient" event class\cite{LAT} for the energies from 100 MeV to 200 GeV.
This event class provides the largest effective area and highest background rates among the LAT standard event classes, which is appropriate for transient sources with small background integration time.
This combination of the GBM and LAT data results in joint spectral fits using forward-folding techniques covering over six decades of energy.

\begin{figure}[tbhp]
\begin{center}
\includegraphics[width=0.8\linewidth]{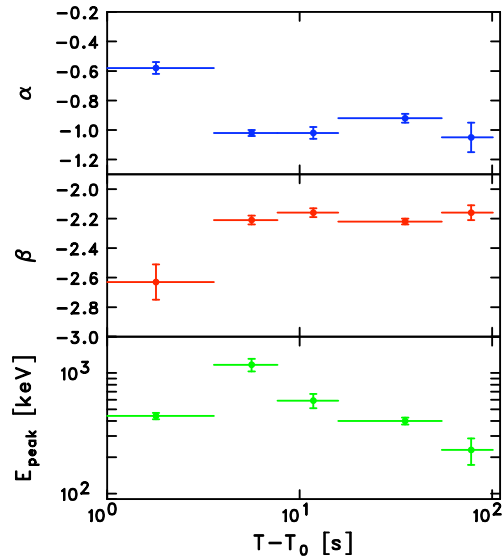}
\caption{
Fit parameters for the Band function, $\alpha$, $\beta$ and $E_{\mathrm peak}$ as a function of time.
}
\label{fig:fit-param}
\end{center}
\end{figure}

\begin{figure}[tbhp]
\begin{center}
\includegraphics[width=0.9\linewidth]{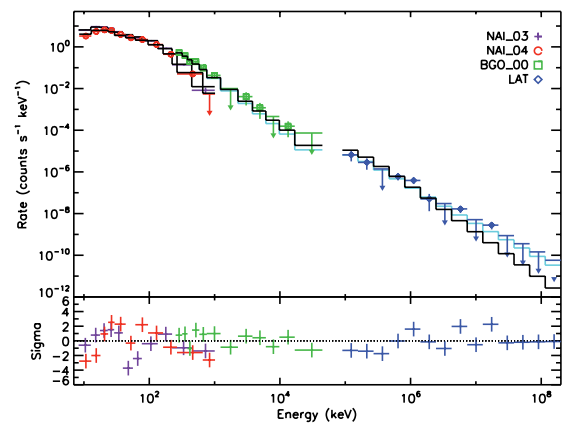}
\caption{
Count spectrum for NaI, BGO and LAT in time bin~(d): the data points have $1\sigma$
error bars while upper-limits are $2\sigma$.
The histograms show the number of counts
%count models 
obtained by folding the photon model
through the instrument response models.
The black histogram shows the count models obtained by folding the best-fit Band photon function
through the detector responses.   
A fit which is a combination of the Band function and a high-energy power law is shown with the cyan histogram.
The residuals are calculated for the simple Band function fit in the bottom panel. 
}
\label{fig:spectrum}
\end{center}
\end{figure}

The spectra of all five time intervals are well fit by the empirical Band function\cite{Band93} which smoothly joins low- and high-energy power laws.
Fig.~\ref{fig:fit-param} shows time evolutions of fit parameters, $\alpha$, $\beta$ and \epeak.
The first time interval, with a relative paucity of photons in the LAT, also has the most distinct spectral parameter values.
The low-energy photon index $\alpha$ is larger (indicating harder emission) and the high-energy photon index $\beta$ is smaller (indicating softer emission) - consistent with the small number of LAT photons observed at this time. 
After the first interval there was no significant evolution in either $\alpha$ or $\beta$. 
In contrast, \epeak, the energy at which the energy emission peaks in the sense of energy per photon energy decade, evolved from the first time bin to reach its highest value in the second time bin, then softened through the remainder of the GRB.
The higher \epeak\ and overall intensity of interval (b), combined with the hard value of $\beta$ that is characteristic of the later intervals, are the spectral characteristics that lead to the emission peaking in the LAT lightcurve (Fig.~\ref{fig:GBM-LAT-LC}). 
%The spectrum of interval (b) with a Band function fit is shown in Fig.~\ref{fig:spectrum}. 
%Comparing the parameters of this interval to the ensemble of EGRET burst detections: the flux at around 1~MeV and $\beta$ are similar to those for GRB 910503 and \epeak\ resembles that for GRB 910814\cite{baring06}.

We searched for deviations from the Band function, such as an additional component at high energies\cite{Gonzalez03}. 
Three photons in the fourth time bin had energies above 6~GeV. 
We tried modeling these high-energy photons with a power law as an additional high-energy spectral component. 
Compared to the null hypothesis that the data originated from a simple Band GRB function, adding the additional power-law component resulted in a probability of 1\% that there was no additional spectral component for this time bin; with five time bins, this is not strong evidence for any additional component.
Fig.~\ref{fig:spectrum} shows the spectrum of this interval with two fit models, a Band function and a combination of the Band function and an additional high-energy power law component.
Our sensitivity to higher-energy photons may be reduced at $z\sim4.35$ through absorption by Extragalactic Background Light (EBL). 
Because the effect of various EBL models ranges widely\cite{razzaque08, ss06}, we do not use EBL absorption effects in our estimation of significance, which gives conservative estimates.

%******************************************************
% Temporarily extended emission
%******************************************************
\section{Long Lived Emission}
Although the lightcurves shown in Fig.~\ref{fig:GBM-LAT-LC} indicate that during interval (e) the spiky structures typical of prompt GRB emission appear to be dying out, the emission persisted in some of the GBM NaI detectors at a low level out to nearly $T_0+200$~s. 
Since the excess above background in the 12 NaI detectors occurred in the ratios expected for the geometry of the detectors relative to the burst direction and this type of low-level, extended emission is a known phenomenon in at least some long GRBs\cite{connaughton02}, we associate it with the GRB and fit the spectrum with a power-law index of $-1.92\pm0.21$. Emission beyond $T_0+200$~s fell below the threshold of the GBM detectors. 

Long lived emission in the LAT band was searched for by a unbinned maximum likelihood fit of a power-law spectral function for a point source at the GROND-determined burst location using the event class with minimum backgrounds (``diffuse'').
The fits are performed for two time intervals $T_0+100$--$200$~s and $T_0+200$--$1400$~s for events within 15\deg\ of the GROND localization coordinates.
The upper bound was chosen because after $T_0+1400$~s, the GRB off-axis angle increased from 50\deg\ to 62\deg\ resulting in decreased effective area.
Contributions from instrumental, Galactic, extragalactic components were included in the fit, as well as the bright source Vela (which is located 13\deg\ from the GRB).
%The most suitable class to study faint sources with minimum backgrounds (``diffuse'') was used to select events within 15\deg\ of the GROND localization coordinates between $T_0+100$~s and $T_0+1400$~s, which were then examined for possible connection with the GRB source. 
%The interval up to $T_0+200$~s was treated separately for correlation with contemporaneous data from the GBM. 
%The upper bound was chosen because after $T_0+1400$~s, the GRB off-axis angle increased from 50\deg\ to 62\deg\ resulting in decreased effective area.
%We performed unbinned maximum likelihood fits of a power-law spectral function for a point source at the GROND-determined burst location in these two time intervals.
%Contributions from instrumental, Galactic, extragalactic components were included in the fit, as well as the bright source Vela (which is located 13\deg\ from the GRB).
Fits in both time intervals show the presence of significant flux. 
For the final time interval, $T_0+200$ to $T_0+1400$~s, the fit yields a flux of $(6.4\pm2.5)\times10^{-6}$ $\gamma\ \mathrm{cm}^{-2}\ \mathrm{s}^{-1}$ for $E > 100$~MeV with a power-law photon index of $-2.8\pm0.5$ at a significance of 5.6$\sigma$. 
If the position of the point source is left free instead of fixed to the GROND localization, the fit yields a source position of RA = $07^h57^m33^s$, Dec. = $-57^\circ00^\prime00^{\prime\prime}$ with an uncertainty of 0.51\deg\ at 90\% C.L. 
This location is 0.45\deg\ from, and in agreement with, the GROND GRB position. 
To solidify the association of this extended emission with the GRB, we performed the same source detection procedure for data from $T_0-900$~s to $T_0$ and no emission was observed. 
A search for emission beyond $T_0+1400$~s was also fruitless.

\begin{figure}[htbp]
\begin{center}
\includegraphics[width=0.9\linewidth]{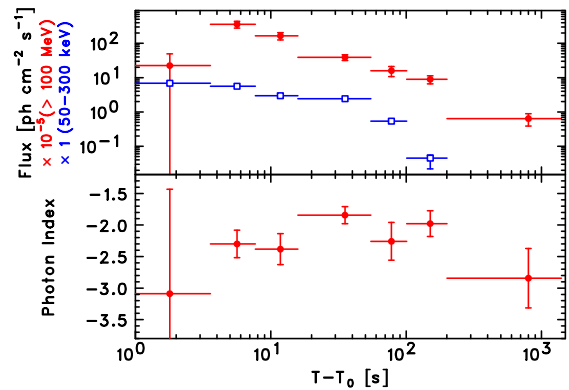}
\caption{
Fluxes (top panel) for the energy range $50-300$~keV (shown in blue open squares) and above 100~MeV (red filled squares), and power-law index as a function of the time from $T_0$ to $T_0+1400$~s (bottom panel, LAT data only).
Red points are obtained by spectral fits of the LAT-only data for all time intervals.
Blue points are obtained with the Band functions listed in Table 1 for the first 5 intervals and a power-law fit with index $-1.90 \pm 0.05$ for the 6th interval.
}
\label{fig:extended}
\end{center}
\end{figure}

We therefore associate this long-lived component with the GRB and include it as a sixth and seventh time interval for comparison with the early-time emission (Fig.~\ref{fig:extended}).
In the LAT data, a constantly declining high-energy flux with a power-law decay index of $-1.2\pm0.2$ is seen throughout $T_0+1400$~s (red points, Fig.~\ref{fig:extended}).
On the other hand, the flux in the GBM band shows a slower decay initially and an apparent break in the lightcurve at $\sim T_0+55$~s. 
The power-law decay indices are approximately $-0.6$ and $-3.3$ before and after the break, respectively. 
%Previous reports\cite{Hurley94,Gonzalez03} have provided tantalizing clues that distinct high-energy components may be a feature of some GRBs.

%******************************************************
% Discussion
%******************************************************
\section{Discussions}
The {\em Fermi} observations of GRB 080916C show that the event energy spectra up to $\sim 100\;$s are consistent with a single model (Band function), suggesting that a single emission mechanism dominates.
A non-thermal synchrotron emission is the favored emission mechanism at keV to MeV energies\cite{meszaros}, and can indeed reach $\sim 30(\Gamma/1000)[5.35/(1+z)]\;$GeV\cite{Peer04}.
However, it should be accompanied by a synchrotron self-Compton (SSC) spectral component produced from electrons that Compton upscatter their synchrotron photons to $\gamma$-ray energies potentially in the LAT energy band.
The apparent absence of an SSC component indicates that the magnetic energy density is much higher than the electron energy density or that the SSC $\nu F_\nu$ spectrum peaks at $\gg 10\;$GeV and thus cannot be detected.
It should be noted that sensitivity to a high-energy additional spectral component is reduced because EBL can absorb high-energy photons via pair-production interactions although the effect of the EBL cannot be estimated reliably due to large model dependence.
%In all models for EBL absorption, the uncertainties at high-$z$ are considerable, so that additional measurements by {\em Fermi} at high-$z$ may be valuable in constraining these models. 

The delayed onset of the GRB 080916C LAT pulse, which coincides with the rise of the second peak in the GBM light curve (see Fig.~\ref{fig:GBM-LAT-LC}) indicates that the two peaks may originate in spatially distinct regions.
A $\gamma\gamma$ pair-production opacity effect may be ruled out since we do not observe any spectral cutoff in the combined GBM/LAT data.
It is intriguing that long lived gamma-ray emission in the LAT band exhibits different temporal behaviors from those in the GBM band.
In particular, our measurement indicates a temporal break in the GBM band in contrary to the continuous decay in the LAT band.
Before our observations, a high-energy (100 MeV -- GeV) tail was observed most clearly from GRB 940217\cite{Hurley94} in observations by EGRET, which was not conclusive.
The LAT high-energy tail may indicate cascades induced by ultrarelativistic ions accelerated in GRBs\cite{Dermer06}, or angle-dependent scattering effects\cite{Wang06}. 

A measurement of redshift $z\approx4.35$ by GROND\cite{greiner09} combined with {\em Fermi} measurements provides several kinematic constraints on this GRB.
Between 10~keV and 10~GeV in the observer's frame, we measure a fluence $f = 2.4\times 10^{-4}$ ergs cm$^{-2}$ which gives at $z\approx4.35$ an apparent isotropic energy release for a standard cold dark matter cosmology with cosmological constant $\Omega_\Lambda= 0.73$, $\Omega_m = 0.27$, and a Hubble's constant of 71 km s$^{-1}$ Mpc$^{-1}$ of $E_\mathrm{iso} \approx 8.8\times 10^{54}$ ergs.  
This is $\sim 4.9$ times the Solar rest energy, and therefore strongly suggests on energetic grounds, for any stellar mass progenitor, that the GRB outflow powering this emission occupied only a small fraction ($\lesssim 10^{-2}$) of the total solid angle, and was collimated into a narrow jet.

Given the intensity of observed photons, large bulk Lorentz factor $\Gamma$ is required to avoid the attenuation of high-energy photons in a compact emission region expected from rapid variability \cite{Krolik91}.
Using the Band function as the target radiation field and setting to unity the optical
depth $\tau_{\gamma\gamma}$ to $\gamma$-ray pair production
attenuation of the highest-energy observed photon, 
we obtain $\Gamma_{min} \approx 608\pm15$ and $887 \pm 21$ in time
bins d and b, respectively, for $z\approx4.35$.
Our limits are much higher than previous firm estimates of $\Gamma_{min} \approx 100$ 
from GRB 990123\cite{Lithwick01}.
The high-energy photons and a large redshift can also provides an limit on the violation of Lorentz invariance expected from some quantum gravity models\cite{amelino98}.
In the linear approximation, the difference in the arrival times $\Delta t$ is proportional to the ratio of photon energy difference to the quantum gravity mass, $\Delta E/M_{\rm QG}$, and depends on the distance the photons traveled. 
The arrival time of the 13.2-GeV photon relative to $T_0$, $t = 16.54$~s, is a conservative upper limit on its $\Delta t$ relative to $\sim {\rm MeV}$ photons, and implies a robust lower limit on the quantum gravity mass, $M_{\rm QG} > 1.30 \times 10^{18}$~GeV$/c^2$, which is an order of magnitude higher than the previous limit obtained in this fashion\cite{albert08}. 
This lower limit is only one order of magnitude smaller than the Planck mass, $1.22 \times
10^{19}$~GeV/$c^2$.

\section{Acknowledgement}
The {\em Fermi} LAT Collaboration acknowledges support from a number of agencies and institutes for both development and the operation of the LAT as well as scientific data analysis. These include NASA and DOE in the United States, CEA/Irfu and IN2P3/CNRS in France, ASI and INFN in Italy, MEXT, KEK, and JAXA in Japan, and the K.~A.~Wallenberg Foundation, the Swedish Research Council and the National Space Board in Sweden. Additional support from INAF in Italy for science analysis during the operations phase is also gratefully acknowledged.
The {\em Fermi} GBM Collaboration acknowledges the support of NASA in the United States and DRL in Germany, and thanks Lisa Gibby, Al English and Fred Kroeger.

%\IEEEtriggeratref{21}
%\bibliographystyle{IEEEtran.bst}
%\bibliography{FermiBIB}   %>>>> bibliography data in report.bib

\end{document}